\documentclass[fleqn,usenatbib]{mnras}

\usepackage[T1]{fontenc}

\DeclareRobustCommand{\VAN}[3]{#2}
\let\VANthebibliography\thebibliography
\def\thebibliography{\DeclareRobustCommand{\VAN}[3]{##3}\VANthebibliography}

\usepackage{graphicx}	
\usepackage{amsmath}	
\usepackage{amssymb}	
\usepackage{newtxtext,newtxmath}

\title[GRB 191016A]{GRB 191016A: A highly collimated gamma-ray burst jet with magnetised energy injection}

\author[M. Shrestha et al.]{M. Shrestha,$^{1}$\thanks{E-mail: m.shrestha@ljmu.ac.uk}
I. A. Steele,$^{1}$
S. Kobayashi ,$^{1}$
N. Jordana-Mitjans, $^{2}$
R. J. Smith,$^{1}$
 H. Jermak,$^{1}$
D. Arnold,$^{1}$
\newauthor C. G. Mundell,$^{2}$
 A. Gomboc, $^{3}$
C. Guidorzi $^{4}$
\\
$^{1}$Astrophysics Research Institute, Liverpool John Moores University, Liverpool Science Park IC2, 146 Brownlow Hill, \\
$^{2}$Department of Physics, University of Bath, Claverton Down, Bath, BA2 7AY, UK\\
$^{3}$Center for Astrophysics and Cosmology, University of Nova Gorica, Vipavska 13, 5000 Nova Gorica, Slovenia \\
$^{4}$Department of Physics and Earth Science, University of Ferrara, via Saragat 1, 44122 Ferrara, Italy
}

\date{Accepted XXX. Received YYY; in original form ZZZ}

\pubyear{2021}

\begin{document}
\label{firstpage}
\pagerange{\pageref{firstpage}--\pageref{lastpage}}
\maketitle

\begin{abstract}
Long gamma-ray burst GRB 191016A was a bright and slow rising burst that was detected by the \textit{Swift} satellite and followed up by ground based Liverpool Telescope (LT).  LT follow-up started $2411$-s after the \textit{Swift} Burst Alert Telescope (BAT) trigger using imager IO:O around the time of the late optical peak.  From $3987-7687$-s, we used the LT polarimeter RINGO3 to make polarimetric and photometric observations of the GRB simultaneously in the $V,R$ and $I$ bands. The combined optical light curve shows an initial late peak followed by a decline until 6147-s, 6087-s, and 5247-s for $I,R$ and $V$ filters respectively followed by a flattening phase.  There is evidence of polarization at all phases including polarization ($P = 14.6 \pm 7.2 \%$) which is coincident with the start of the flattening phase.  The combination of the light curve morphology and polarization measurement favours an energy injection scenario where slower magnetised ejecta from the central engine catches up with the decelerating blast wave.  We calculate the minimum energy injection to be $\Delta E / E>0.36$.  At a later time combining the optical light curve from BOOTES (reported via GCN) and IO:O we see evidence of a jet break with jet opening angle $2\degr$.
\end{abstract}




\section{Introduction}
Gamma-ray bursts (GRBs) are extremely energetic explosive events happening at a cosmological distances. GRBs can be broadly divided into two groups based on gamma-ray duration as short GRB (<2 s) and long GRB (>2 s). The short GRBs are thought to be produced by the merger of compact objects such as binary neutron stars or a neutron star and a black hole. Whereas, long GRBs are thought to be produced by the explosion of very massive stars. In both cases, a collimated relativistic jet is launched which produces prompt emission as $\gamma$ rays from internal dissipation and interaction of the jet with local ambient medium produces afterglow in various wavelengths \citep{Piran_1999,Zhang_2004}. 

Dedicated instruments such as \textit{Swift} \citep{Gehrels_2004} have increased the number of GRB detections and through advancements in robotic capabilities of ground based instruments, we can do a rapid follow-up at various wavelengths.  Studies of long GRBs show that there is considerable diversity within the group \citep{Panaitescu_2008,Kann_2010,Panaitescu_2011,Zaninoni_2013,Wang_2018,Wang_2020}. \citet{Panaitescu_2008} examined 30 GRBs and found the optical light-curve of GRB afterglows could be divided into four different categories 1) fast-rising with an early peak, 2) slow-rising with a late peak, 3) flat plateaus, and 4) rapid decays. In their sample, 5 GRBs are considered to be slow-rising and all have optical peak time greater than 500 seconds. They attribute these variations in the optical light-curve to offset in the viewing angle of the observer. However, \citet{Panaitescu_2011} found 17 GRB to have plateau in the light-curve and proposed that the plateaus could be due to the energy injection.
In contrast \citet{Oates_2009} found in their analysis of 27 Ultraviolet Optical Telescope (UVOT) optical/ultraviolet light-curves of GRBs, the significant fraction of the sources rises in the first 500 seconds and all of them decay after 500 s. This behaviour is explained by the start of the forward shock according to \citet{Oates_2009}. 

\citet{Kumar_2000} showed that when slower ejecta from the central engine interacts with faster ejecta that has been slowed down due to interaction with ambient medium, energy could be injected to forward shock/blast wave. This can also create a flat plateau in the light curve. During this interaction, a reverse shock in the ejecta could be produced which tend to carry large scale magnetic fields \citep{Steele_2009, Mundell_2013}. Thus, in this scenario, a plateau phase can be seen potentially along with a high level of polarization. In this paper, we present a study of a long, slow rising GRB with a photometric plateau phase and detections of polarization which provides evidence for this scenario.  

\subsection{GRB 191016A}

\begin{figure*}
\centering
\includegraphics[width=1.8\columnwidth]{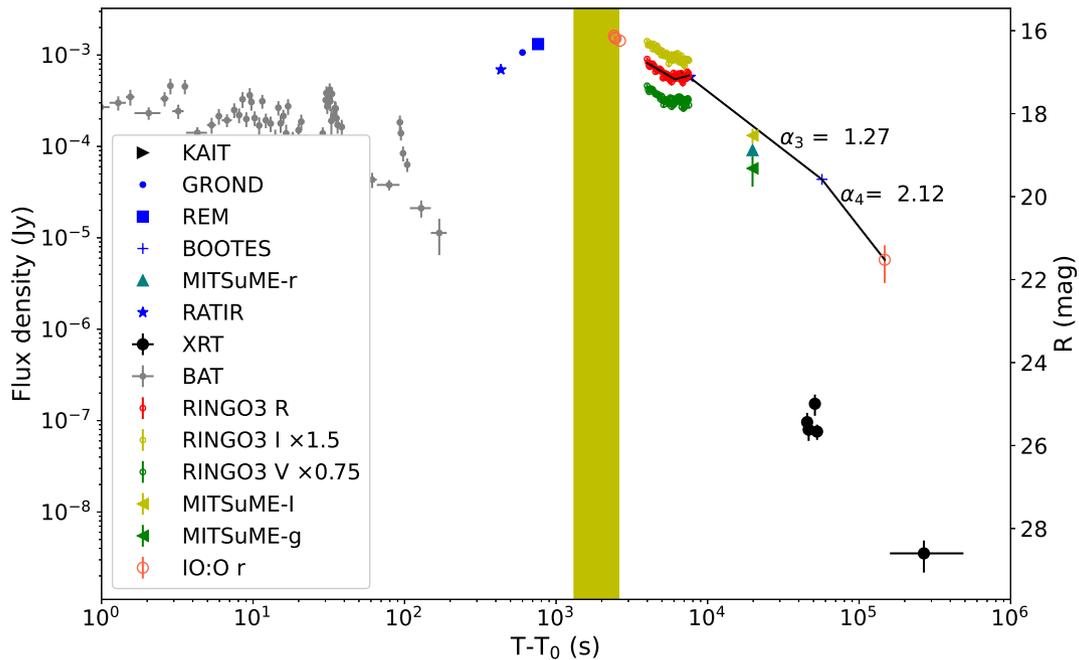}
\caption{Light curve of BAT and \textit{Swift} XRT data from \citet{Evans_2009} along with optical data from ground based follow-up in blue as reported in GCN circulars. We present BAT data binned to signal-to noise ratio of 5, and the XRT data are for observed flux at 1 keV. RINGO3 $R$ and IO:O $r$ data are presented in red and the lower limit of MITSuME R data along with I and g data is presented. Broken power law fit is shown for the end of the RINGO3 $R$ data, BOOTEs, and IO:O late time data. The decay index are $1.27 \pm 0.14$ before the break and $2.12 \pm 0.5$ after the break. The shaded region is the time interval for possible optical peak from TESS observation \citep{Smith_2021}.} 
\label{fig:XRT-BAT}
\end{figure*}

This paper presents results of optical photometric and polarimetric analysis of the {\textit Swift} BAT transient GRB 191016A
\citep{GCN26008}. GRB 191016A was triggered by \textit{Swift} Burst Alert Telescope (BAT) at 04:09:00 UT \citep{GCN26008}. Due to Moon observing constraints \textit{Swift} didn't slew to the BAT position until 15:17 UT, thus there is no \textit{Swift} XRT and UVOT data for early times.  The BAT and XRT light curves are shown in Fig.~\ref{fig:XRT-BAT} which is taken from light-curve repository of \textit{Swift} \citep{Evans_2009}. BAT data analysis at $15-350$ keV showed two overlapping pulses with two peaks occurring at $\sim$ $T_0-10$-s and $T_0+35$-s with $T_{90}$ = 220 $\pm$ 180-s \citep{Barthelmy_2019}. A Simple power-law fit to the BAT data gives photon index of 1.541 $\pm$ 0.095 \citep{Barthelmy_2019}.  A power-law fit to XRT light-curve gives a temporal decay index of $1.9^{+0.5}_{-0.3}$ and a photon index of $1.9^{+0.6}_{-0.4}$ \citep{Page_2019}.

The GRB was followed-up by various  facilities as recorded in GCN \footnote{https://gcn.gsfc.nasa.gov/} (e.g. COATLI \citep{Watson_2019}, KAIT \citep{Zheng_2019}, RATIR \citep{Watson_2019}, BOOTES \citep{Hu_2019}, MITSuME \citep{Toma_2019}, REM \citep{Melandri_2019}, GROND \citep{Schady_2019}). The photometric redshift of the burst was estimated at 3.29 \citep{Smith_2021}.  The Transiting Exoplanet Survey Satellite (TESS) \citep{Ricker_2015} was monitoring the same portion of the sky as GRB 191016A and was able to make a serendipitous observation of the rise and fall of the optical afterglow. The brightest peak of the TESS light-curve was estimated to occur between $1316$-s and $2590$-s post-trigger \citep{Smith_2021} (shaded region in Fig.~\ref{fig:XRT-BAT}) where the large uncertainty is due to the 30 minute TESS full frame observing cadence. In any case, the optical counterpart is late-peaking from both the ground and space based data. 

In this paper, our data set consists of a small quantity of optical photometry taken around the peak time plus a much larger quantity of photometry and polarimetry over the time interval 3987-s to 7587-s post-trigger at which time the optical counterpart was bright and initially slowly declining.  We find a plateau in the light-curve starting at $6146$-s,$6087$-s, and $5247$-s for $I$, $R$, and $V$ band and detection of polarization at all epochs with a high ($P=14.6\pm7.2\%$) value around the time of the start of the plateau for $I$ band data. In addition, we use late time observational data from IO:O, BOOTES, and MITSuME to demonstrate a broken power-law behaviour at late times.  

The paper is organized as follows: in Section~\ref{sec:obs} we describe the instruments and observations, in Section~\ref{sec:reduction} we describe the photometric data reduction process. In Section~\ref{subsec:reduction-polarimetry}, we present the polarization reduction process and discuss the technique to determine polarization detection and interstellar polarization. In Section~\ref{sec:discussions}, we provide implications of photometric and polarimetric analysis and finally provide conclusions in Section~\ref{sec:conclusions}.

\section{Observations}\label{sec:obs}

Data from two different Liverpool Telescope \citep{Steele_2004} instruments (IO:O and RINGO3) are presented in this paper.  LT is a 2.0 meter fully robotic telescope at Observatorio del Roque de los Muchachos, La Palma \citep{Steele_2004}. The telescope is optimized for time-domain astrophysics and efficient for rapid automated follow-up of rapidly varying sources such as GRBs \citep{Guidorzi_2006}. IO:O \footnote{https://telescope.livjm.ac.uk/TelInst/Inst/IOO/} is a conventional CCD imager equipped with $u'g'r'i'z'$ filters and a $10\times10$ arcmin field of view. In this paper, we present results for the $r'$ filter.  

RINGO3 \citep{Arnold_2012} was a fast-readout optical imaging polarimeter that was online at LT from early 2013 to January 2020. It had a field of view of $4 \times 4$ arcmin and used a wire grid polarizer that rotated at $\sim 0.4 Hz$. It simultaneously observed polarized images in three different wavebands using three separate electron multiplying CCDs. The three wavebands very roughly correspond to the standard astronomical $I$, $R$ and $V$ bands with effective wavelength of $8500 $ \AA, $7050 $ \AA, and $5300 $ \AA  ~respectively.  Exposures are synchronised with the phase of the polarizer's rotation and each camera receives eight exposures per rotation. Analysis of these eight images provides the different linear Stokes vectors. RINGO3 produces 24 CCD frames (8 per camera) every 2.3 seconds which for this paper have been stacked into 60-second blocks for photometry and 600-second blocks for the polarimetry.

There was a delay in LT observations of the GRB because it was BAT only burst and the software in LT did not initially slew correctly to the GRB. $r'$ band IO:O observations, therefore, began 40 minutes after the trigger time. RINGO3 observations then commenced 66 minutes after the trigger time in the $I$, $R$, and $V$ filters. Even though RINGO3 observations commenced after $\sim1$ hour, the afterglow was still bright enough to be easily detected in all the filters.  

\section{Photometric Analysis}\label{sec:reduction}

\subsection{Data Reduction}\label{subsec:reduction-Photometry}

\begin{figure*}
\centering
\includegraphics[width=1.95\columnwidth]{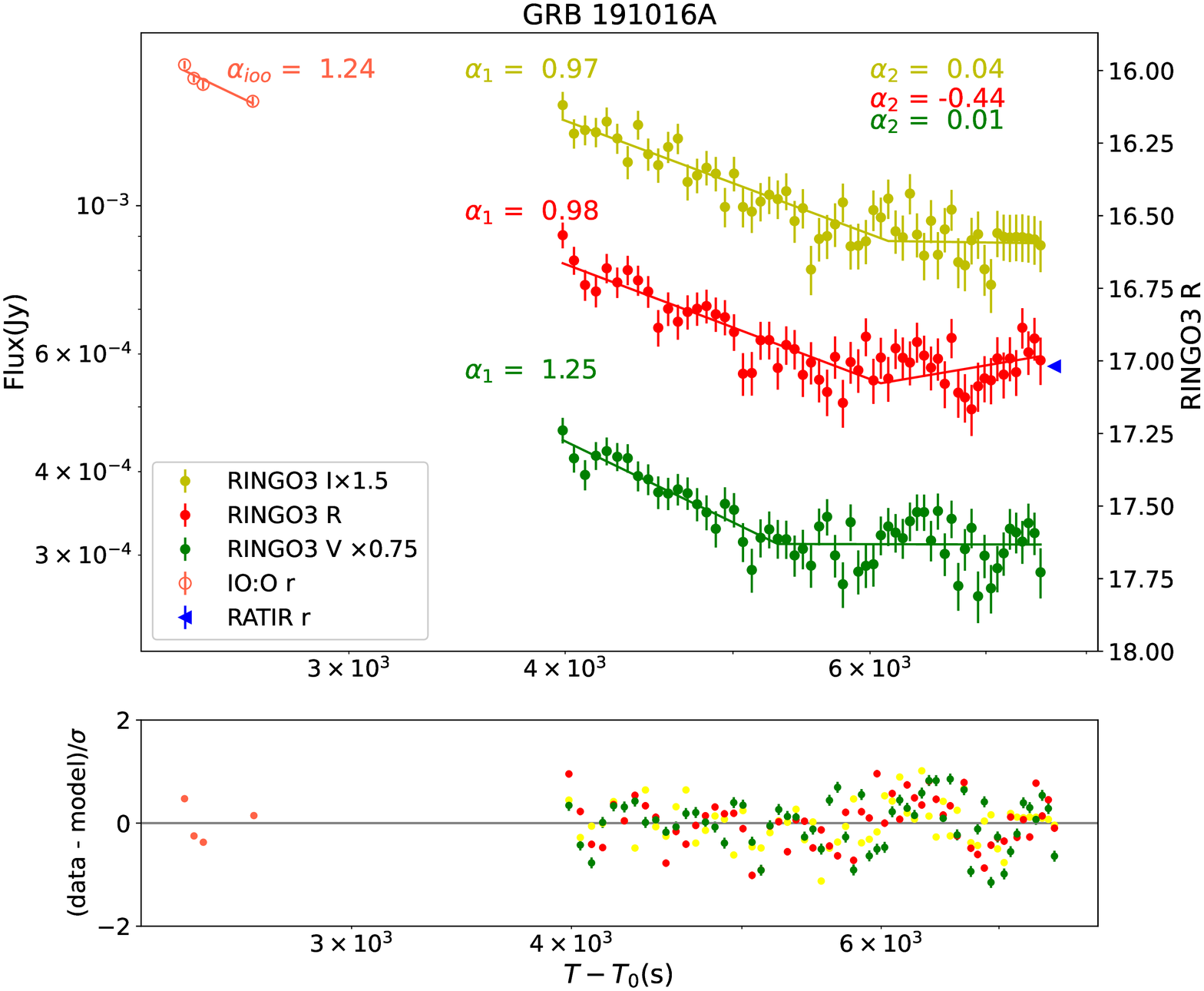}
\caption{(top) Light curves of the GRB 191016A observed by RINGO3 in three different bands and IO:O in r' band. Broken power-law fit are performed on these data. Temporal power-law decay index ($\alpha_1$) before the break time of 6146 s, 6087 s, and 5247 s for $I$, $R$, and $V$ filters respectively and ($\alpha_2$) after the break time is presented in the plots for three different bands for RINGO3 data. IO:O data are plotted with a power-law fit and decay index value is given. The blue triangle indicates the RATIR $r$ data point \citep{Watson_2019} used for cross calibration. Flux density given in left y-axis is converted to RINGO3 $R$ magnitude. (bottom) Comparison of broken power-law fit to the data is provided for different time period. The reduced $\chi^2$ for these fits are 1.04, 1.01, and 0.9 and degrees of freedom are 56, 56, and 56 for $V$, $R$, and $I$ bands respectively. }
\label{fig:photometry_results}
\end{figure*}

 Basic CCD reduction for both IO:O and RINGO3 instruments uses a common pipeline which includes bias subtraction, dark subtraction, flat fielding and World Coordinate System fitting \footnote{https://telescope.livjm.ac.uk/TelInst/Pipelines/}.  To extract source counts, we used the AstroPy {\sc Photutils} package \citep{Bradley_2019}.
 The IO:O $r'$ data was calibrated with respect to a Johnson-Cousins like $R$ band \citep{Bessell_1998} using USNO-B.1 \citep{Monet_2003} field stars and converted to a flux density using the zero points from \citet{Bessell_1998}.  

RINGO3 photometry was carried out by stacking all 8 polaroid rotation angles into a single image per 60 seconds epoch.  Since the RINGO3 photometric bands are non-standard considerable colour terms affect the photometry \citep{kopac_2015}.  Because the other stars in the frame of this particular afterglow have a very limited colour range it was not possible to solve for this directly.  Instead we applied the technique outlined in \cite{Jermak_2017} and \cite{Jordana_2020}.  This technique calibrates the natural photometric system of the instrument.  It uses observations of unreddened A0 stars that by definition must have zero colour index in the Vega magnitude system for all combinations of filters. These can be combined with the wavelength dependent throughput of RINGO3 \citep{Arnold_2017,Jordana_2020} to calculate flux densities at the effective wavelengths of the RINGO3 filters.

Since no observations of A0 targets were made on the same night as the GRB, we instead used the mean observed zeropoints in archive observations of all such targets taken with RINGO3 in the same year of observation to calibrate our data. As the typical properties of the variable component of extinction from La Palma extinction are similar at all wavelengths \citep{Stickland}, this procedure should yield a reasonable estimate of the spectral energy distribution even though the absolute calibration may be incorrect.  To make an absolute calibration we, therefore, applied a multiplicative factor to the R band counts (equivalent to an additive offset in magnitude) to ensure that our light-curve overlapped with simultaneous $R$ band data obtained by the KAIT \citep{Zheng_2019} and RATIR \citep{Watson_2019} experiments (see Figure \ref{fig:photometry_results}).  Since no overlapping data were available in the $V$ and $I$ bands we applied the same correction factor as applied to the $R$ band data.  

All data points (from our data and other experiments plotted in the figure) have also been corrected for Galactic and host galactic extinction as per \citet{Schlegel_1998} using $5\degr \times 5\degr$ statistics which produces E(B-V) of 0.1047 $\pm 0.0025$ and yields $A_V =0.329 $, $A_R =0.276$, $A_I = 0.206 $ \footnote{https://irsa.ipac.caltech.edu/cgi-bin/bgTools/nph-bgExec} for Galactic extinction and host galactic extinction are $A_V =0.354 $, $A_R =0.285$, $A_I = 0.207 $ from \citet{Smith_2021}  . 

The resulting three band light curves are presented in Figure \ref{fig:photometry_results} and show an initial decline followed by a plateau phase that appears to have a correlated structure between the bands. Fitting \footnote{All errors on derived quantities and fit parameters quoted as $\pm$ in this work are 1$\sigma$ equivalent.} shows that all three optical light curves are best fitted with a broken power-law (BPL) which produces lower reduced $\chi^2$  compared to a simple power-law (PL). In addition to the reduced $\chi^2$ test, we also performed an F-test to compare the BPL vs PL fit and found that the BPL was a better fit in all three cases with $F>38$ and $p<0.0011$ (Table~\ref{tab:photometry}.) For broken power law, we let the break time to be a free parameter and find the best fit for different filter data separately. We also performed a simultaneous BPL fit to all three different wavelength data and found an initial decay index of 0.9 before a break time of 5487-s and decay index of -0.28 after the break. The reduced  $\chi^2$ compared to individual fit is much higher of 9.25, indicating the slight differences in the fits between the three bands are likely real.
The fitting parameters for both BPL and PL models such as  different decay indices ($\alpha$) and the $1\sigma$ errors, reduced $\chi^2$ values, break time, p-value, and F-value are presented in Table~\ref{tab:photometry}. \footnote{The absolute calibration uncertainty of the RINGO3 data does not affect the fit parameters.}

\begin{figure*}
\centering
\includegraphics[width=1.95\columnwidth]{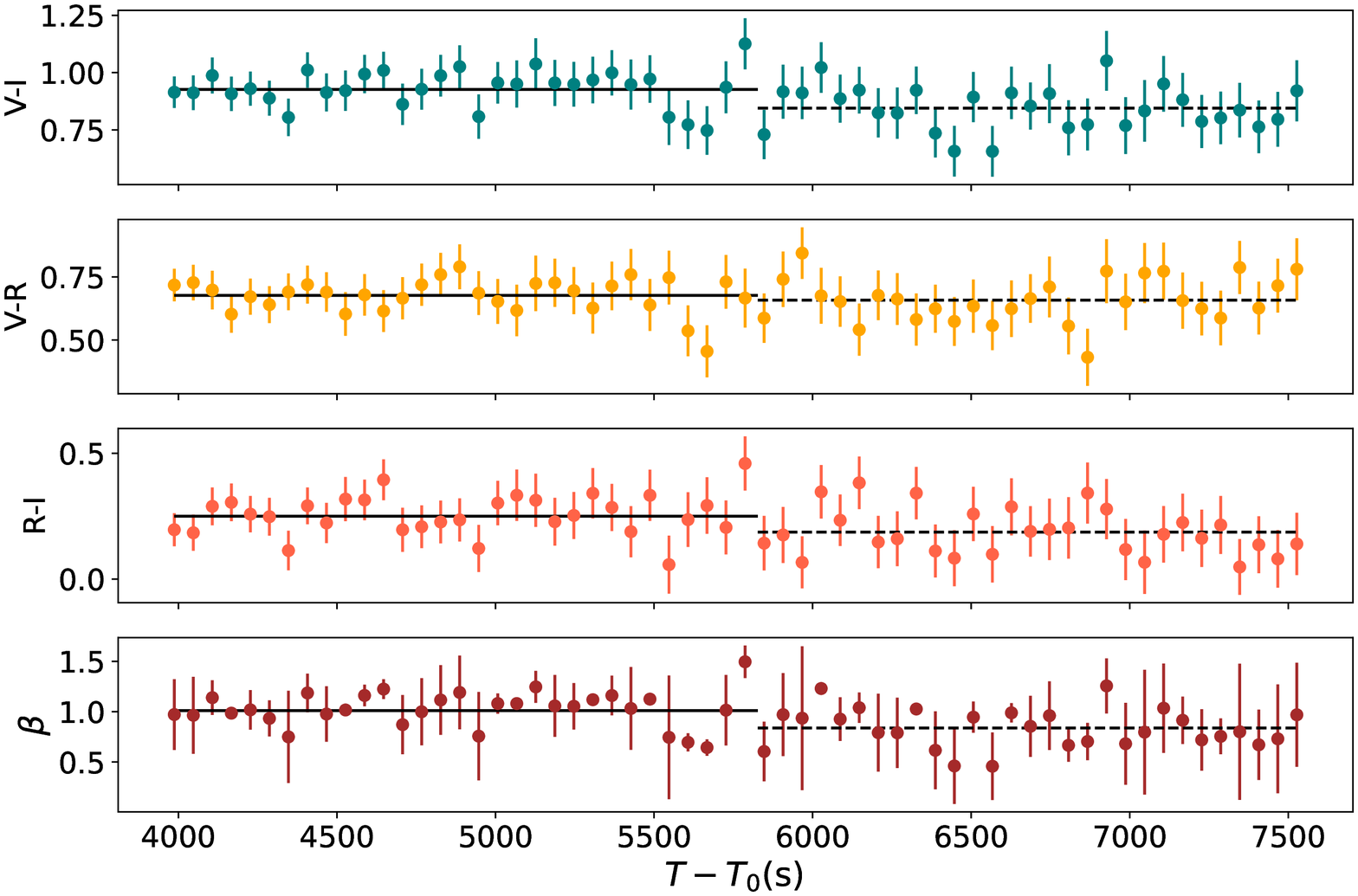}
\caption{Colour index with respect to time for the top three plots. Top plot is for V-I, middle plot is for V-R, and the bottom plot is R-I. Solid black line is the mean of values before the break time of 5827 s  which is the average of three break times and the black dashed line is the mean of data points after the break time. Mean before the break are $0.93 \pm 0.013, 0.67 \pm 0.012,$ and $0.26 \pm 0.013$  and mean after the break are $0.85 \pm 0.017, 0.647 \pm 0.016,$ and $0.203 \pm 0.017$ for V-I, V-R, and R-I respectively. Bottom plot is the spectral index $\beta$ with respect to time with a solid black line is the mean value ($1.05 \pm 0.03$) before the break and the dashed line is for the mean ($0.80 \pm 0.04$) after the break, where errors are standard error. }
\label{fig:color}
\end{figure*}

In Figure \ref{fig:color}, we plot the time evolution of the RINGO3 $V-R$, $R-I$ and $V-I$ colours as a function of time.  A significant colour change to the blue  is seen for the case of $V-I$ at the start of the plateau phase. For the other colours we see a smaller change as would be expected due to their shorter wavelength baseline. We find $V-I$ initially is $0.93\pm 0.01$ mag before the plateau phase and changes to $0.81\pm0.02$ during the plateau phase. Similarly $V-R$ and $R-I$ change from $0.67\pm 0.01$ , $0.26\pm0.01$ to $0.65\pm0.02, 0.20\pm0.02$ respectively from initial to plateau phase.    Using the RINGO3 effective wavelengths the resulting spectral energy distributions were well fit with a single power law of index $\beta$ with mean value $1.05\pm0.03$ before the plateau and $0.85\pm0.04$ during the plateau (Fig. \ref{fig:color}). 

\section{Polarization Analysis}\label{subsec:reduction-polarimetry}

\subsection{Data Reduction}

For polarization analysis, we treat the eight RINGO3 frames separately.  We used the same initial data reduction and count extraction procedure as outlined in Section \ref{subsec:reduction-Photometry}.  To extract polarization of the source we implemented the procedure presented by \citet{clarke_2002}, which is utilised for RINGO, RINGO2, and previous RINGO3 analysis (e.g. \citet{Jermak_2016}). This technique provides linear Stokes parameters $q$ and $u$. Correction for instrumental polarization is accomplished by the use of regular observations of non-polarized standard stars \citep{Schmidt_1992} and subtraction of the instrumental $q$ and $u$ values (Table \ref{tab:inst}) from the measured values of the science target (see \citet{Jermak_2016} for more details of this procedure). An analysis of a small sample of unpolarized standards taken around in the weeks leading up to the GRB was used to calculate these values.  \cite{Jermak_2017} showed that no correction for instrumental depolarization was necessary for RINGO3 data.

To calculate the electric vector position angle we rotate the measured the position angle ($\psi$) based on the telescope Cassegrain axis sky position angle (SKYPA), measured east of north:
\begin{equation}
    EVPA = \psi + SKYPA + \Theta.
    \label{eq:evpa}
\end{equation}
Here $\Theta$ is a calibration factor that gives the position angle offset combined of the angles between the orientation of the polarizer, the telescope focal plane, and the trigger position of the angle measuring sensor. It was calculated using polarized standard observations done by the instrument, the values of $\Theta$ are provided in Table~\ref{tab:inst}.

\begin{figure*}
\centering
\includegraphics[width=1.95\columnwidth]{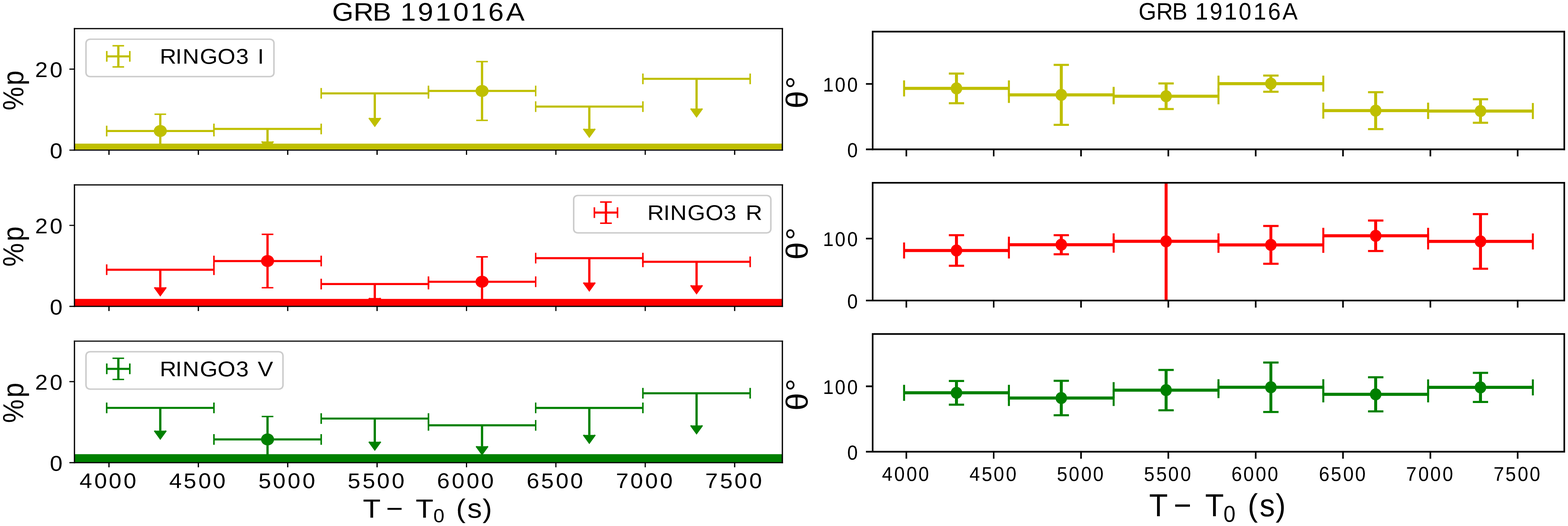}
\caption{Observed polarization degree with respect to $T-T_0$ (left) . The results in I, R, and V bands are presented in the top, middle, and bottom panels respectively. Upper limits and detections are presented accordingly. Error on the x-axis is the observation time for the given data point. Position angle plot is presented in the right panel. The shaded region is the sum of host galactic and Galactic dust contribution to polarization.  }
\label{fig:pol_results}
\end{figure*}

Finally, we calculate the error\footnote{Upper limits on polarization are shown for 1$\sigma$} in polarization degree (\textit{p}) and EVPA. Since \textit{p} is always positive, there is a polarization bias introduced by noise in \textit{q} and \textit{u}. We correct for this and calculate the polarization error using the prescription developed by \cite{Plaszczynski_2014}. 
To calculate the error in position angle we used standard error propagation theory. 
The results of our analysis are presented in Table \ref{tab:polarization} and plotted in Figure \ref{fig:pol_results}.

\begin{table*}
    \centering
    \begin{tabular}{c|c|c|c|c|c|c|c|c}
          \hline
           \hline
         Wavelength & Model &$T_{break}$ (s) & $\alpha_1$  & $\alpha_2$ & $\chi_{r}^2 (d.o.f)$ & p-value & F-value\\
         
          \hline
         
          I  & BPL & 6147 & 0.97 $\pm$ 0.07 &  0.04 $\pm$ 0.17&   0.9  (56)& 
          0.447 & 38.12\\
          I  & PL & - & 0.76 $\pm$ 0.05 &  -&   1.2  (57) & 0.0011 & 38.12\\
          R  & BPL&6087 & 0.98 $\pm$ 0.07 &  -0.44 $\pm$ 0.17&   1.01  (56) &0.447 & 38.12\\
          R  & PL&- & 0.63 $\pm$ 0.05 &  - &   1.7  (57) &  0.0011 & 38.12\\
          V  &BPL& 5247 & 1.25 $\pm$ 0.1 &  0.01 $\pm$ 0.09&   1.04  (56) & 0.396 & 48.173\\
           V  &PL& - & 0.57 $\pm$ 0.04 &  -&   1.9  (57) & 0.00005 & 48.173\\
          \hline
    \end{tabular}
    \caption{Results of broken power-law (BPL) and power-law (PL) fitted to the light curve. For different wavelengths, $T_{break}$ where the light-curve is broken, $\alpha_1$ best fit temporal decay index before the $T_{break}$, $\alpha_2$ is the best fit temporal decay index after the $T_{break}$, reduced $\chi_{r}^2$ values and degree of freedom (d.o.f) for the best fit, and the p-value for this $\chi^2$ values, and F-value comparing the BPL and PL is presented in the last column.} 
    \label{tab:photometry}
\end{table*}

\begin{table*}
    \centering
    \begin{tabular}{c|c|c|c|c|c|c}
          \hline
           \hline
          Waveband & $q_{in}$ & $\sigma$  & $u_{in}$ & $\sigma$ & $\Theta (\degr$) & $\Theta (\sigma$) \\
          \hline
          V & -0.005 $\pm$ 0.008& 0.033  &  -0.018 $\pm$ 0.016 &0.066   & 85.5 & 6.5 \\
         R & -0.012 $\pm$ 0.004&0.019   &  -0.035 $\pm$ 0.004& 0.015  & 82.8 & 6.39 \\
         I & -0.014 $\pm$ 0.006& 0.025  &  -0.035 $\pm$ 0.004& 0.018  & 84.5 & 5.9 \\
         \hline
         \hline
         
    \end{tabular}
    \caption{Table of instrument $q$, $u$, and $\Theta$ factor values for the time periods closer to GRB 191016A observations using RINGO3. We quote standard error ($\frac{\sigma}{\sqrt{N}}$) where N is the total number of observations as the error in instrument $q$ and $u$. Standard deviation ($\sigma$) is also presented. These values are much smaller than error in Stokes $q$ and $u$ of the GRB 191016A. }
    \label{tab:inst}
\end{table*}

\begin{table*}
    \centering
    \begin{tabular}{c|c|c|c|c|c|c|c|c|c|c}
     \hline
      \hline
         GRB & t-$t_0$ (s) & P (\%) (I) & P (\%) (R) & P (\%) (V) & EVPA (deg) & Rank\\ 
         \hline
         191016A & 3987-4587& $\bf 4.7\pm4.1$ & $<9.1$ &  $< 13.4 $ & $93.16 \pm 22.6$ (I)& {\bf 0.99 (I)}, 0.63 (R), { 0.91} (V)\\
         191016A & 4587-5187&$<5.2$ &$\bf 11.2\pm6.6$ & $\bf 5.7\pm5.6$ & $90.1 \pm 15.4$ (R), $82 \pm 26.4$ (V) & 0.59 (I), {\bf 0.99} (R), {\bf 0.98} (V) \\
         191016A & 5187-5787&<14.0 & $<5.5$& <10.8&- & 0.72 (I), 0.11 (R), 0.72 (V) \\
         191016A & 5787-6387& $\bf 14.6\pm7.2$ & $\bf 6.1\pm6.0$ & <9.2 & $100 \pm12.4$ (I), $89.8 \pm 30.5$ (R) &{\bf 0.99} (I), {\bf 0.95} (R), 0.73 (V) \\
         191016A & 6387-6987& <10.7 & <12.0& <13.5&-& 0.67 (I), 0.83 (R), 0.86 (V)\\
         
         191016A & 6987-7587&<17.6 & <11.0& $< 16.8 $& - & 0.76 (I), 0.86 (R), {0.94} (V) \\
         
          \hline
          \hline
    \end{tabular}
    \caption{Columns are GRB identifier, time range, polarization degree for I, R, and V bands, Position angle, permutation rank for detected polarization.  Detections that pass both error bar and permutation (rank) analysis are highlighted in bold.  EVPAs are only given in the case of a positive detections. }
    \label{tab:polarization}
\end{table*}

    
        

We note that the moon was nearly ($\sim 95$\%) full and close ($\sim 20^\circ$) to the source on the sky.  In order to confirm that the resulting polarization of the sky background was not affecting our measurements
 we measured polarization values of several other sources in the field of view. For objects both brighter and fainter than the GRB, we measured polarization < 3\% in our data.  Therefore, even though the moon was bright and the sky quite strongly polarized, we see no evidence of significant sky contamination in our GRB measurements.

\subsection{Detection Confirmation}\label{sec:detection}

Table \ref{tab:polarization} identifies 5 possible detections in GRB 191016A (highlighted in bold) at various epochs and wavelengths from consideration of the error bars (where an object that has an error bar that does not cross zero is identified as a possible detection).  Given we have made multiple measurements, it is therefore important to consider the possibility that these results may be false-positives.  To illustrate this, consider a simple case where measurements with error bars that do not cross the y axis can be viewed as a ``$2\sigma$'' detection.   Each measurement would therefore have an associated probability of being a false positive of 5\% (i.e. 1 in 20) if the error distribution was Gaussian.  Since we made 18 measurements of something that has a 1 in 20 chance of occurring, we might therefore reasonably expect that the chance of one of the measurements showing positive would be very high.  However having 5 out of 18 positives seems unlikely (the the exact probability can be calculated using the binomial theorem as shown below).

To make a more robust analysis along these lines, we need to know the true probability of each detection.  We can calculate this using the permutation analysis technique described in detail in \cite{steele-ringo2-2017}.  Briefly this technique shuffles the measured counts from the 8 rotor positions into all possible permutations and evaluates the implied polarization for each permutation.  Because shuffling the data should remove the coherent signal associated with a true detection but leaves the stochastic noise characteristics unaffected, it provides a method of testing the probability of the measured polarization being consistent with a null hypothesis of no polarization for an individual observation.

Since this method relies on measured counts it will also be sensitive to instrumental polarization.  In order to correct for this we assume that the bright stars in the field of the GRB 191016A to have low polarization.  This assumption was checked to be reasonable by comparison with the counts ratios for zero polarized standards taken around the same time.  Taking the ratio of the source (GRB) and the brighter  star counts for each rotor position then removes instrumental polarization \citep{steele_2006}. After correcting for this instrumental effect, we created $(8-1)!$ permutations of the corrected count for the GRB source to generate 5040 different sets of eight flux values. We use these sets of eight flux values to calculate the polarization degree and create a rank which tells us the probability of the detected polarization degree. We can then check null hypothesis; if the source is unpolarized then what is the chance of getting some polarization signal due to noise in the data? For example if the rank is greater than 0.95 it means the probability of being an unpolarized source $p = 1-rank$ will be $<0.05$. 

The rank values for each epoch and waveband are presented in Table \ref{tab:polarization}.  This analysis identifies all 5 of our possible detections as having $p<0.05$.  If we assume a null hypothesis that the GRB was unpolarized at all epochs then the cumulative binomial probability of 5 out of 18 $p<0.05$ observations is $p<0.002$ for it to have arisen by chance.  Overall we are therefore confident that the source is truly polarized.  As an aside, we note that 17 of the 18 observations have $p<0.5$, corresponding to a cumulative binomial probability of $p<0.00007$ and indicating that polarization below the individual observation's detection limit was likely present at all times during the RINGO3 observing period.

We note there is a perfect match between the observations that passed the standard error bar analysis and those that passed the rank test at $p<0.05$.
While formally looking at the individual detection for the rank analysis alone, the probability of one of the five being a false positive is $\sim0.6$ and of two being false positives $\sim0.2$, this one-to-one mapping between the two methods means that overall we are confident that they are all in fact likely to be true detections of polarization.  

For $I$-band data, we see clear detections for the first, and fourth data points, with polarization degree values of $4.7\%$ and $14.6\%$ and ranks 0.99, and 0.996 respectively. Thus, we see that the probability of this being a zero polarized source with random noise is $p = 1-rank$ which is 0.01, and 0.004 or $1\%$ and $0.4\%$ respectively. Thus, we class the measured polarization as a detection. For $R$-band we see clear detections for the second, and fourth data points and via the permutation method we find a rank of 0.99, and 0.95, with a measured polarization degree of $11.2\%$ and $6.1\%$ respectively. For $V$-band we get detection for the second data point with rank values of 0.98. These data point has $2\%$ probability of being a zero polarized source with polarization signal due to noise. The detected polarization degree value is  $5.7\%$.  We note that the EVPA values for all the filters are consistent with time; the median values are $82.2 \degr$, $92.7 \degr$, and $92.1 \degr$ for $I$, $R$ and $V$ filters respectively.

\subsection{Interstellar and Host Galaxy Polarization} \label{ISP}

We use galactic extinction values  from \citep{Schlegel_1998} to estimate the Galactic interstellar polarization (GISP) using the formulation by \citet{serkowski_1975}. For the co-ordinates of GRB 191016A, the value of $E_{B-V}$  is 0.1047 \citep{Schlegel_1998}. First we used $p^V(GISP) \leq 9 E_{B-V}$ to calculate the upper limit in polarization induced by GISP in the V band. After this calculation, we used $p/p_{max} =\exp[-K {\rm ln}^2(\lambda_{max}/\lambda)]$ to calculate GISP in the $R$ and $I$ bands using V band as the $\lambda_{max}$ and $p^V$ as the $p_{max}$; where $p$ is the polarization induced by GISP at the wavelength $\lambda$, $p_{max}$ is the maximum polarization induced by GISP at the wavelength $\lambda_{max}$, and $K$ is a constant given to be 1.15 in \citet{serkowski_1975} and later modified to be $K = -0.10+1.86 \lambda_{max}$ by \citet{Wilking_1982}.  
The resulting GISP estimates for this GRB are $0.77\%$, $0.87\%$, and $0.94\%$ in $I$, $R$, and $V$ bands.  GISP will therefore make a negligible contribution to our polarization measurements and we do not attempt to correct for it.

We also estimated the ISP contribution of the host galaxy. From multi-band photometry \citet{Smith_2021} finds a best fit model with $A_V = 0.35$ for the host galaxy. Calculating the ISP using this extinction value for Milky Way like dust gives $0.78\%$, $0.88\%$, and $0.95\%$ for $I$, $R$, and $V$ bands. However, \citet{Smith_2021} shows the best fit model is Small Magellanic Cloud (SMC) like dust for the host galaxy and using SMC like dust we estimate ISP to be $0.6\%$, $0.75\%$, and $0.95\%$ for $I$, $R$, and $V$ bands respectively.  This is again negligible compared to our measurement limits (see Fig. \ref{fig:pol_results}).


%



\section{Discussion} \label{sec:discussions}

In this section we present a coherent model that attempts to explain the key observational features of GRB~191016A in a coherent fashion:
\begin{itemize}
    \item A late optical light curve peak occurring between 15 and 40 minutes after the initial explosion
    \item An initial optical light curve decline from 40 minutes to 90 minutes with power-law index $\alpha \sim 1.2$
    \item An optical light curve plateau extending from 90 minutes to at least 125 minutes and a corresponding increase in optical polarization during that phase.
    \item A post plateau optical and X-ray light curve that shows evidence for a jet break at around 52800 seconds.
\end{itemize}


\subsection{Late Peak}

As shown in Fig. \ref{fig:XRT-BAT} the rising pre-peak light curve of the afterglow was detected by a number of facilities and the combination of TESS and other data constrains it to occurr at $15 \mbox{mins} \lesssim T-T_0 \lesssim 40 \mbox{mins}$.  This time can be understood as the onset of 
afterglow.  Considering that this is a relatively high redshift event $z=3.29$,  this late 
onset can be understood to be due to the cosmological time dilation.  Combining the onset time with the standard time dilation correction gives a bulk Lorentz factor  $\Gamma = 90-130$ from \citet{Smith_2021} which is typical for the class. The lower range of bulk Lorentz factor (90) assumes $E_{iso} = 2.37 \times 10^{53}$ ergs, $t_{peak} = 2590 s$ and they assume the particle density of surrounding medium and radiative efficiency to be 1. In the case of upper limit of bulk Lorentz factor, they assume $E_{iso} = 6.61 \times 10^{53}$ ergs, $t_{peak} = 1316 s$ and they assume the particle density of surrounding medium and radiative efficiency to be 1.
As an aside, we note the decaying high energy emission at early times ($T-T_0<$ a few hundred seconds) detected by Swift-BAT can be attributed to the high latitude emission of the internal dissipation process \citep{Zhang_2006}.  

The late peak could also be explained by the offset of viewing angle as theoretically shown by \citet{Rossi_2002}. In few cases of late peaking GRBs such as GRB 081028 and GRB 080710, the late peak in the optical light curve is attributed to off-axis viewing angle \citep{Margutti_2010, Kruehler_2014}. This is also one of the possible explanations presented for GRB 191016A by \citet{Smith_2021}. 

\subsection{Initial Decline}

The pre-plateau emission can be well understood as a standard forward shock dominated jet, originating from a blast wave with a constant explosion energy $E$.  The synchrotron radiation is likely to be in the slow cooling regime when our observations were carried out. Depending on the emission frequency the two possible closure relations
are $\alpha=3\beta/2$ for $\nu_m < \nu < \nu_c$ and $\alpha=(3\beta-1)/2$
for $\nu > \nu_c$. Simple substitution shows that our observed pre-plateau emission 
characteristics $\alpha \sim 0.97-1.23$ and $\beta \sim 1.05$ correspond better with the regime where  $\nu > \nu_c$.  In this case the electron energy distribution index is $p=2\beta \sim 2.1$.

In all three bands, we detect polarization ($4.7 \pm 4.1$\%, $11.2 \pm 6.6$\%, and $5.7 \pm 5.6$\%) during the initial decay phase which is forward shock dominated. For GRB 091208B, \citet{Uehara_2012} reported optical polarization of $10.4 \pm 2.5 \%$ during forward shock phase with temporal decay index of $0.75 \pm 0.02$. They attribute the polarization signature to magnetohydrodynamic instabilities which in turn causes position angle to fluctuate with time. For GRB 191016A, the position angle is not fluctuating with time as shown in Fig.~\ref{fig:pol_results} (right). Thus, this scenario is not favourable for GRB 191016A. \citet{Jordana_2021} reported detection of low level of polarization ($2.8^{+2.0}_{-1.6}$ in the early forward shock from GRB 141220A which they attribute to dust in the host galaxy.  This is not the case for GRB 191016A as the host galactic dust contribution and GISP is low (Fig.~\ref{fig:pol_results}(left)).

Another scenario where forward shock emission could be polarized is when the line of sight runs almost along the jet edge. Such a viewing angle happens by chance only for narrow jets ($ 1/\Gamma \sim 10^{-2}$ \citep{Gruzinov_1999}) which is consistent with the bulk Lorentz factor from the timing of the late peak. To produce polarized emission, the tangled magnetic fields generated in the forward shock need to be anisotropic in the shocked ambient comoving frame \citep{Ghisellini_1999,Sari_1999}, on top of the favoured viewing angle. The magnetic fields parallel and perpendicular to the shock normal can in principle have different averaged strengths because the shock normal is a special direction at any point in the shock front. For this time period and viewing angle very close to the jet, we get polarization prediction of $\sim 5\%$ from \citet{Sari_1999} considering a jet break to be around BOOTES observation time of 56880 s (more detail analysis about this jet break time in Section~\ref{sec:post_p}). This predicted polarization value is within the error bars of our detected polarization. In addition, this viewing scenario has implications for light curves as shown by \citet{Rossi_2002}. As described above the late peak in optical light-curve could be explained by this offset in viewing angle. Finally we note that the jet break is predicted to be smoother for an off-axis viewing angle by \citet{Ghisellini_1999}. However, due to our relative lack of data in late time, we cannot say either way regarding the jet break feature.  Overall this is our most favoured explanation for the detected polarization during this relatively early phase.

\subsection{Plateau Phase}

The most striking feature of the optical light curve is the flattening at $T-T_0\sim 5.8 \times 10^3$ sec combined with the high degree of polarization at that time.  This flattening can be  
interpreted by two possible models.  The first model is energy injection to the forward shock/blast wave. This energy injection would happen when slower ejecta from the central engine catches up with the decelerating blast wave \citep{Kumar_2000}. The second model requires density enhancements in the ambient medium.  In this model the blast wave either collide into them or surfs on them, depending on the density contrast \citep{Lazzati_2002}. However, 
since magnetic fields generated within shocks are highly tangled,  the synchrotron 
emission from the blast wave is not expected to be polarised.  The
simultaneous detection of the polarization signals, therefore, favours the energy injection model. 
Our polarization measurements of other events indicate that ejecta from the central engine often carries large scale magnetic fields \citep{steele-ringo2-2017}. The collision between the ejecta and the blast wave can cause a reverse shock in the ejecta.
The transient reverse shock emission might therefore be expected to induce the observed higher polarization signals around the flattening time. We also note the flux above the cooling frequency does not depend on the ambient density, the fact we are in the spectral regime $\nu > \nu_c$ is also evidence in favour of the energy injection model. 

We now assume that the kinetic energy of the ejecta $\Delta E$ is given to the blast wave after their collision. The blast wave makes a transition from a Blandford-McKee solution \citep{Blandford_1976} with energy $E$ to another with energy $E+\Delta E$. Since the afterglow emission above the cooling frequency depends on the energy as $F_{\nu} \propto E^{(p+2)/4}$, the emission should become brighter by a factor of $[(E+\Delta E)/E]^{(p+2)/4}$ than the extrapolation of the original decay line. Once the blast wave settles into the blast wave with $E+\Delta E$ the afterglow would then be expected to start to decay again with the original decay index $\alpha$.  Although we do not observe this phase of the light curve we can obtain a lower limit of $\Delta E$, by assuming that the afterglow starts to decay again at our last data point.  We find that compared to extrapolation of the pre-plateau flux, the post-break fitted lines are brighter by a factor of $1.39$ (the averaged value over the three bands) at $t\sim 7600$s. We therefore determine $\Delta E/E\gtrsim 0.36$.

When the slow ejecta catches up with the decelerating blast wave, the ejecta and the blast wave have similar Lorentz factors and the collision is mildly relativistic \citep{Kumar_2000}.  At the shock crossing time, the forward- and reverse- shocked regions have the same Lorentz factor and internal energy density. Using analyses similar to those in \citet{Kobayashi_2003} and \citet{Zhang_2003}, we can obtain approximated relations at the reverse shock crossing time,

\begin{equation}
  \label{eq:t}
  \begin{aligned}
   & \nu_{m,r}/\nu_{m,f} \sim R_B \Gamma_\times^{-2},  \\
 & \nu_{c,r}/\nu_{c,f} \sim R_B^{-3},  \\
& F_{\nu,max,r}/F_{\nu, max, f} \sim  R_B \Gamma_\times (\Delta E/E),
  \end{aligned}
\end{equation}

where the subscripts $f$ and $r$ indicate forward and reverse shock, respectively, $F_{\nu,max}$ is the peak flux in the spectral domain, $R_B=(\epsilon_{B,r}/\epsilon_{B,f})^{1/2}$ is introduced to take into account the higher magnetisation of the ejecta, $\Gamma_\times$ is the Lorentz factor of the shock regions at the shock crossing time $t_\times \sim 7000$ sec. Since the cooling frequency of the reverse shock is lower than that of the forward shock for $R_B > 1$,  assuming the optical band is above both cooling frequencies, we have the flux ratio of the two shock emissions at the optical band $F_{\nu,r}/F_{\nu,f} \sim \Delta E/E$ for $p\sim 2$. Using our constraint $\Delta E/E\gtrsim 0.36$ and assuming the reverse shock is highly polarized (as was found for example in \citep{Steele_2009, Mundell_2013}, we would therefore expect, as observed, high measured optical polarization signals ($14.6 \pm 7.2$ in $I$ band and $6.1 \pm 6.0$ in $R$ band).

 In addition to reverse shock emission, we could also get some polarization degree due to the viewing angle of the observer. For this time period compared to the time of the jet break, the polarization degree is expected to be closer to $0 \% $ for $q = 0.95$ case from \citet{Sari_1999}, whereas for $q = 0.71$ case we can get polarization signal closer to our detection, where $q$ is the ratio of the offset of the angle of the line of sight of the observer to the centre of the jet to the jet's initial angular size \citep{Sari_1999}. According to jet break models \citep{Sari_1999,Ghisellini_1999}, the light curve is expected to decay as a regular forward shock emission $t^{-1}$ before the break, and then decay faster $t^{-p}$. We need an additional mechanism (e.g. energy injection) to explain the flattening phase. If there is an energy injection phase, the deceleration of the forward shock becomes less significant or equivalently the observable region with angular size $1/\gamma$ around the line of sight grows slower. The peak in the polarization curve for $q=0.71$ can broaden and we can get the polarization degrees closer to our observation. However, most of the observations so far near the jet break as shown by \citet{Covino_2016} are lower than our detected polarization. It indicates that magnetic fields generated in shocks are not as highly anisotropic as \citet{Sari_1999} and \citet{Ghisellini_1999} assume. Thus, short-lived reverse shock is the most likely scenario for this detected polarization within the error bar.  Once the reverse shock crosses the ejecta, a rarefaction wave starts to propagate from the inner edge of the eject to the blast wave, the reverse shock energy is quickly transferred to the blast wave with a timescale comparable to the shock crossing time \citep{Kobayashi_2007}.  Since no electrons are accelerated in a reverse shock after the shock crossing, the reverse shock emission above the cooling frequency and the polarization signals are expected to quickly disappear after the shock crossing. 

It is more challenging to explain the colour change during the plateau phase.  An interesting possibility is that the electron energy distribution indexes are different in the blast wave and the reverse shock because the reverse shock is sub-relativistic (i.e. the ejecta and the blast wave have similar Lorentz factors at the collision).  The theoretical predictions from diffusive (Fermi) shock acceleration are $p = 2$ at non-relativistic shock speeds and $p \sim 2.22$ at ultra-relativistic velocities \citep{Sironi_2015}.  The electron energy distribution in the reverse shock region might have a smaller $p$. However, shock acceleration theories still predict $p> 2$ or $\beta >1$ in our case, whereas we find $\beta\sim0.8$ at that time.

In another scenario, if a dense clump in the surrounding medium contains large scale, ordered magnetic fields, the forward shock emission can be polarized when the shock propagates through the clump \citet{Granot_2003}. The clump may be due to the wind ejection history of the progenitor star or associated with a supernova remnant.  However, since the optical band is above the cooling frequency of the forward shock, the density enhancement in the ambient medium does not affect the temporal decay index as long as the enhancement is moderate and the forward shock surfs on them (e.g. \citet{Lazzati_2002,Nakar_2007}). Thus, this mechanism is not suitable to explain the optical light curve flattening which happens simultaneously with the detection of the polarization signals.

\subsection{Post Plateau Phase}\label{sec:post_p}

The late X-ray light curve shows a steep decay $\alpha\sim 1.9$ (+0.5, -0.3), and 
our IO:O data along with BOOTES data \citep{Hu_2019} also suggest a steep decay after the RINGO3 observations to BOOTES data of $\alpha\sim 1.27\pm0.15$ whereas from BOOTES to late IO:O data shows $\alpha\sim 2.12\pm0.15$  (Fig \ref{fig:XRT-BAT}). There is an upper limit value reported in GCN for MITSuME (\citep{Toma_2019}) $R_C$ band. We ignored this data for the fit as it is an upper limit instead of detection and the effective wavelength is different from RINGO3, BOOTES, and IO:O $r'$ band filters.  However, the $I_C$ band of MITSuME is closer to the $R$ band of RINGO3 and BOOTES, and the $r'$ band IO:O filter.  Thus, we also checked the BPL fit including $I_C$ data from MITSuME with our $R$ band data.  In this case, the BPL fit shows an initial decay of $1.39 \pm 0.12$ before the break time of 56880 s and after the break, the decay index is $1.89 \pm 0.53 $. \footnote{ The MITSuME \citep{Toma_2013} team have also reported g' band detection in the GCN, we have included the data point in the Fig.~\ref{fig:XRT-BAT}, however, due to lack of any other late time data in a similar band, we can not use it.}
Regardless of the exact nature of the BPL index changes, they can be explained if a jet break occurs between the RINGO3 and IO:O observations around the time of the BOOTES observation.  Because of the difference in filters we use the BPL fit with only BOOTES and late time IO:O data in our analysis below.

This interpretation is also supported by the XRT data (Fig. \ref{fig:XRT-BAT}; since a jet break is 
dynamical or/and geometrical effect,  the break should be monochromatic. The conventional jet break model predicts a temporal decay index of $\alpha = p\sim 2.1$ after the jet break \citep{Rhoads_1999,Sari_1999b}.



Although the jet break time depends on viewing angle as well, we can use the standard formula to obtain a rough estimate of jet opening angle $\theta_j$ which is given by:
\begin{multline*}
    \theta_j \sim 2.0 \left(\frac{t_b}{12579 \mbox{s}}\right)^{3/8} \left(\frac{1+z}{4.29}\right)^{-3/8}
\left(\frac{E_{iso}}{1.19\times10^{54}\mbox{erg}}\right)^{-1/8} \\
\left(\frac{n}{0.1\mbox{cm}^{-3}}\right)^{1/8}  \mbox{deg}.
\end{multline*}
For $E_{iso}$ we assume conversion efficiency to be 0.2, thus $E_{iso} = 2.37 \times 10^{53} / 0.2 $ where $2.37 \times 10^{53}$ erg is from \citet{Smith_2021}. For ISM density $n = 0.1 cm^{-3}$, we find the jet opening angle to be $2\degr$ which shows that the jet is highly collimated. Although the isotropic energy ($1.19 \times 10^{54}$ erg) is close to the solar mass energy, the geometrically corrected value ($7.24 \times 10^{50}$ erg) is relatively small in the long GRB population \citep{Frail_2001}. 

\section{Conclusions} \label{sec:conclusions}
We have analyzed GRB 191016A follow-up observational data from the Liverpool Telescope fast readout polarimeter RINGO3 and imager IO:O combined with data from GCN circulars.  We presented photometric light curves and polarization measurements in the $V$, $R$ and $I$ bands. At later times light-curve, we utilized GCN values from BOOTES \citep{Hu_2019} along with late IO:O observations.  From our observations we concluded:

\begin{itemize}
    \item The light curve in all three bands can be best fitted by broken power-law. The initial decay phase can be explained as forward shock model with temporal decay index of $0.97 \pm 0.07$, $0.98 \pm 0.07$, and $1.24 \pm 0.1$ in $I$, $R$, and $V$ bands. After an initial break at $6147 \pm 30$ s, $6087 \pm 30$ s, and $5247 \pm 30$ s for $I$, $R$, and $V$ filters respectively the light-curve is flat with a decay index close to 0. This can be interpreted as a period of energy injection.
    
    \item Polarimetry was also done in three bands simultaneous with the photometric observations.  At 5 out of 18 waveband-epochs we found significant ($p<0.05$) polarization detection.  The strongest ($P=14.6\pm7.2$\%) polarization detection with a $99.6\%$ confidence level was made in the $I$ band at the start of the plateau phase. This polarization detection along with a flat light curve at that epoch favours an energy injection scenario over the enhancement of the ambient medium.  This is because in the energy injection case we expect the signature of a transient reverse shock which can contain an ordered magnetic field producing a short-lived polarized emission. Our light-curve analysis shows an energy injection of $>36\%$ compared to the initial energy.
    
    \item \textit{Swift} XRT late time data (45390-s after BAT trigger) shows steep decay with temporal decay index of 1.9 (+0.5, -0.3) and late time (56880-s) optical decay using BOOTES \citep{Hu_2019} and IO:O data shows a similarly steep decay of 2.12 $\pm 0.2$ which points towards a probable jet break with the caveat that there are few data points. Using these assumptions, we calculated the jet opening angle to be $2\degr$.
\end{itemize}

Our analysis demonstrates how combining time-resolved polarimetry with multi-wavelength photometry is a powerful diagnostic tool in determining the emission mechanism and the behaviour of the driving central engine in GRBs. Currently, such analysis is generally limited to the brighter bursts, however 
early follow-up of GRBs with more sensitive instruments such as MOPTOP \citep{Shrestha_2020} and future 4-m class robotic telescopes \citep{Jermak_2020} will enable such analyses to be expanded to the much wider sample of bursts crucial to driving our further understanding of GRB physics.

\section*{Acknowledgements}
Operation of LT on the island of La Palma by Liverpool John Moores University at the Spanish Observatorio del Roque de los Muchachos of the Instituto de Astrofisica de Canarias is financially supported by the UK Science and Technologies Facilities Council (STFC).  MS is supported by an STFC consolidated grant to LJMU.  This research made use of Photutils, an Astropy package for detection and photometry of astronomical sources (\cite{Bradley_2019}). AG acknowledges the financial support from the Slovenian Research Agency (research core funding P1-0031, infrastructure program I0-0033, project grants J1-8136, J1-2460) and networking support by the COST Actions CA16104 GWverse and CA16214 PHAROS. CGM and NJM thank Hiroko and Jim Sherwin for financial support. This work made use of data supplied by the UK Swift Science Data Centre at the University of Leicester. We would like to thank our anonymous referee for thoughtful comments that have greatly improved the paper. 

\section*{Data Availability}
All the observational data are freely available online in the LT archive.



\bibliographystyle{mnras}
\bibliography{grb_catalog} 








\bsp	
\label{lastpage}
\end{document}